4

# The Tactical Optimal Strategy Game(TOSG) Protocol Cockpit Software Control For Massive Ordnance Penetrator Release

Dr. Carol A. Niznik
NW SYSTEMS
Rochester, NY
dr_carol_niznik@yahoo.com
**Abstract.** The Massive Ordnance Penetrator(MOP) has been developed to destroy deeply buried nuclear components by controlled release from a B2 or B52 airplane. This type of release must be cockpit software controlled by the Tactical Optimal Strategy Game(TOSG) Protocol to optimally determine the war game aspects of the dueling from other countries' MOP releases, and the depth at which the MOP explosions can occur for maximal safety and risk concerns. The TOSG Protocol characteristics of games of strategy, games of optimal strategy and tactical games are defined initially by the game of strategy as a certain series of events, each of which must have a finite number of distinct results. The outcome of a game of strategy, in some cases, depends on chance. All other events depend on the free decision of the players. A game has a solution if there exist two strategies, which become optimal strategies when each mathematically attains the value of the game. The TOSG Protocol war game tactical problem for a class of games can be mathematically modeled as a combat between two airplanes, each carrying a MOP as the specification of the accuracy of the firing machinery and the total amount of ammunition that each plane carries. This silent duel occurs, because each MOP bomber is unable to determine the number of times its opponent has missed. The TOSG Protocol realizes a game theory solution of the tactical optimal strategy game utilizing the theory of games of timing, games of pursuit, games of time lag, games of sequence, games of maneuvering, games of search, games of positioning and games of aiming and evasion. The geometric software structure for the TOSG Protocol is a game tree identifying the possible depth of explosions. This finite game tree with a distinguished vertex is embedded in an oriented plane to facilitate the definition of a strategy as a geometric model of the character of a game for the successive presentation of alternatives. The tactical optimal strategy determination by the TOSG Protocol Cockpit Software is mandatory for the execution of the correct and maximally effective MOP release by the MOP bomber.

I INTRODUCTION

The Massive Ordnance Penetrator(MOP) GBU-57A/B[8] is the most powerful bunker buster, a 13,000 kilogram weapon, developed by the United States to be used against tunnel imbedded nuclear devices. The MOP can smash through 20 meters of reinforced concrete before exploding. The MOP uses a Hard Target Smart Fuse [14] that allows detonation inside buried or reinforced concrete targets. The detonation occurs by the fuse after a sensor informs the fuse that a weapon has passed through a number of layers or voids in the target. The MOP will be fired by a B2 or B52 Airplane[8].

The unique formal TOSG(Tactical Optimal Strategy Game) Protocol Cockpit Software Control for MOP Release is mathematically defined to assist the MOP bomber's actions. The TOSG Protocol is comprised of three topological models; Model 1 two airplanes with MOP explosives as a Two Person Zero-Sum game [18,23], Model 2 airplane MOP gunner aiming and evasion game[10,11] to fire the MOP at a tunnel, and Model 3 the games of timing to predict timing of the MOP airplane gunner firing[26]. A class of games which are tactical will represent a contest between two players who are trying to obtain the same objective. This tactical game has a solution if there exists two strategies, which are optimal strategies[23] if the integration of their derivative times a function equals the values of the game. The game of strategy consists of a certain series of events, each of which must have a finite number of distinct results[29]. The resource allocation optimization for the TOSG Protocol Management is based on the Theory of Games of Timing to achieve an optimal strategy and optimal timing interval. The optimized process follows a timing chart or message sequence of timed events that must have the timing to be optimized for each process. The first optimal strategy, within the Games of Timing Kernel[7,26]equation achieves the TOSG Protocol optimal strategy and optimal timing intervals and the second optimal strategy makes an Optimal Strategy Decision from the TOSG Protocol Risk Optimization equation within the Kernel of the Games of Timing equation.

II. TOSG PROTOCOL WAR GAME PROCESS
1. Tactical Game Zero-Sum Two-Person Game(MODEL 1)

474

The Tactical Game[23] definition for the TOSG Protocol begins with a Zero-Sum Two-Person Game [18]. Zero-Sum means that the gain of one player is matched by the loss of the other player. In a game the following three observations are given for each player: (a) certain choices available, (b) knowledge of consequences of the choices for each choice of opponent, and (c) choice must be made independent of knowledge of the opponent's decision. There is then a single payoff function, because of only one strategy for each player. The game is defined by the triplet $\Gamma = (X, Y, \Psi)$ where X and Y are two closed sets and $\Psi$ is a real valued, measurable function defined on $X \times Y$; $\Psi$ is called the payoff or utility function. The elements $x \epsilon X$ and $y \epsilon Y$ are called pure strategies; the positive measures with total measure 1 defined over X and Y are called mixed strategies. The game has a solution if there exist two strategies F(x) and G(y) such that,

$$\int \Psi(x,y) \, dF(x) \geq V, \text{ all } y \epsilon Y \quad (1)$$
$$\int \Psi(x,y) \, dG(y) \geq V, \text{ all } x \epsilon X \quad (2)$$

F and G are Optimal Strategies and V is the value of the game. Each game will represent a contest between two players each trying to obtain the same objective. When one of the players succeeds, it will win one unit; the opponent loses the same amount, and the contest is over[23]. Each player has limited resources and can make only a fixed number of attempts to reach the goal. These attempts must be made during the interval $0 \leq t \leq 1$, and each attempt may fail or succeed. At t = 0 every attempt fails; at t = 1 every attempt succeeds. At any other time an attempt made by player 1 will be successful with probability P(t), and will fail with probability 1- P(t). An attempt by player 2 succeeds with probability Q(t) and fails with probability 1- Q(t). The functions P(t) and Q(t) increase continuously. Each player knows these functions and the total number of attempts that its opponent can make.

After the contest begins each player is unable to determine how many unsuccessful attempts have been made by the opponent. This specialize form of combat or War Game between two airplanes P and Q, each carrying MOPs describes the accuracy of the firing machinery and the initial resources related to the total amount of ammunition that each player can carry, i.e., the B2 Airplane currently can carry 2 MOPs, and the B52 Airplane can currently carry 6 MOPs[8]. This problem is often called a silent duel, because it is assumed that each pilot is unable to find out how many times the opponent has fired and has missed[23]. In the formal description of the game x and y will be vectors that describe the times when the attempts will be made and $\Psi(x,y)$ will be the expected gain for player 1.

1.A. Gain For Two Person Zero Sum Game Players[18]

The gain $\Psi(x,y)$ for the MOP bomber[18,23] is stated when it uses strategy x and the opponent nuclear device in the tunnel uses strategy y, ($x \epsilon X, y \epsilon Y$). Then because of the Zero Sum definition, $\Psi(x,y) < 0$ (MOP bomber loses). $\Psi(x,y) =$ gain or loss to the nuclear device in the tunnel opponent. The use of mixed or optimal strategy could also be useful for a more exacting development of traits if the single payoff function did not yield the expected result. One of the traits would then enforce the development of how each pure strategy could be optimally used. The further formalism for mixed strategies redefines the parameters as;

X = MOP bomber m pure strategies = times when attempts are made to fire the MOP = probability vector $\sigma(X)$ э there exists, $\sigma(x_1), \ldots, \sigma(x_m)$,

Y = opponent nuclear weapon in tunnel n pure strategies = probability vector $\tau(Y)$ э there exists, $\tau(y_1), \ldots \tau(y_n)$,

$$\psi(\sigma, \tau) = \sum_{x \epsilon X} \sum_{y \epsilon Y} \sigma(x) \tau(y) \psi(x, y) \quad (3)$$

Then, the Game Solution for mixed strategies $\sigma_o$ and $\tau_o$ э v = $\psi(\sigma_o, \tau_o)$, $\psi(\sigma_o, \tau) \geq \psi(\sigma_o, \tau_o) \geq \psi(\sigma, \tau_o)$ for all $\sigma \epsilon X, \tau \epsilon Y$. The most optimal plan of action for each of the two players is to try to maximize their respective payoff functions. Since the Zero Sum Two Person Sum Game is assumed, the most positive outcome for the MOP bomber must be determined with the following steps[18],

(1) The MOP bomber's opponent tries to minimize its average gain, so the bomber is assured of $\min_{x \epsilon X} \psi(x,y)$,

(2) The MOP bomber's choice of action must be made such that its payoff will be at least, $\max_{x \epsilon X} \min_{y \epsilon Y} \psi(x,y)$,

(3) The payoff to the opponent is the negative of the MOP bomber's payoff, so for any pure strategy y that the opponent choses,

$$\min_{x \epsilon X}(-\psi(x,y)) = -\max_{x \epsilon X}(\psi(x,y)) \quad (4)$$

(4) The MOP bomber can then obtain a payoff of at least $\max_{x \epsilon X} \min_{y \epsilon Y} \psi(x,y)$ or no more than $\min_{y \epsilon Y} \max_{x \epsilon X} \psi(x,y)$, where,

$$\max_{x \epsilon X} \min_{y \epsilon Y} \psi(x,y) \leq \min_{y \epsilon Y} \max_{x \epsilon X} \psi(x,y) \quad (5)$$

The modeling of the MOP bomber's experiences by the game theory formulation will enable the MOP bomber to visualize the risk issues as a consequence of judgments based on experience rather than as an issue composed of a maximum certainty of personal loss. An N Person Game model[18] is also possible for the MOP bomber in reference to other problems, if there are multiple players. However from the basic Two Person Zero Sum Game model the baseline for the majority of situations has been provided.

2. TOSG Protocol Cockpit Software Aiming And Evasion Game Theory (MODEL 2)

The aiming and evasion game theory[10,11] uses a gunner, marksman, or MOP bomber aboard the B2 Airplane or the B52 Airplane firing the MOP at the tunnel location, with a time lag in the MOP gunner's target position in the tunnel. This game theory formulation realizes that the nuclear device will be moved deeper into the tunnel, how and when the marksman should make its prediction and the hit probability. The marksman has an ideal strategy with the property that every near optimal strategy is close to it. One of the most classic of military problems is how best to aim at a mobile target, which is deliberately maneuvering to confound prediction of the position, i.e., the burying deeper of the nuclear weapons in the tunnel. How best will the target maneuver the crucial features in common, i.e. the time lag between detection of the target and arrival of the projectile. The time lag could be composed of , (1) delay between detection of the target and aiming of the firing device, and (2) the flight time of the projectile itself. The lag will be time lag as a whole, a mixed strategy. When the player of a game employs a mixed strategy, it means the decision is not made in accordance with any predetermined certain plan, but involves a certain amount of randomness.

A game theoretic solution prescribes, but does not dictate behavior, and the exact probabilities to respectively minimize or maximize the probability of a hit. If the target were to follow any prescribed certain plan, it would plainly be a ruinous policy as soon as the MOP gunner became aware of it. Any fixed policy of the MOP gunner would enable the target always to escape once the MOP gunner learned the policy[11]. The goal should be for an optimal mixed strategy or policies of best regulated randomness for each player. A sunken nuclear device is aware of an enemy MOP bomber's presence, but the airplane is too high for precise detection. The tunnel nuclear device is interested only in not being hit. The device has no offensive means. The airplane has one or more MOP's and to avoid extraneous factors it is assumed the MOP bombers aim is perfect. The nuclear device in the tunnel knows nothing about when or where the MOP will be dropped after detonation. .The nuclear device can only maneuver to be deposited deeper to minimize the hit probability. The only kinematic restriction is that the nuclear device travels with a fixed speed of v. There is a time lag T between the MOP bomber's last information about the depth of the nuclear device in the tunnel and the detonation. Thus the MOP bomber must aim at an anticipated depth of the nuclear device in the tunnel.

3. TOSG Protocol Game Tree Cockpit Software Structure

The game theory attempts to answer three questions: (1) the optimal strategy of player 1, i.e. the continued changing of the depth of the nuclear device in the tunnel, (2) the optimal strategy of player 2, i.e. when and where should the MOP bomber strike, (3) the value of the game, i.e. what is the hit probability when both players use the best tactics. Therefore, the TOSG Protocol Geometric Cockpit Software Structure realizing these questions is the TOSG Protocol Game Tree[6,15] in Figure 1. The enduring strategy established for the Game Tree realizes that each decision depends on the prior moves. This Game Tree will ensure that the MOP Bomber(P) achieves a near optimal strategy. This Game Tree Model is based on a game theory model [10,11] where there was a bomber aiming at a target that was a battleship. The 1,2,3 are vertical positions in the TOSG Protocol Models that represent positions of the nuclear device(E) in the tunnel. The probabilities of E reaching each of the vertical positions 1, 2, 3, each deeper in the tunnel are: $(1-x)^2$, x , x(1-x). According to game theory P will elect the largest of these three probabilities for the calculation of the detonation location for the firing of the MOP. The best possible x for E is the value that renders the maximum of the three polynomials a minimum at V, the value of the game, a root of $x = (1-x)^2$, i.e., V = .382. Therefore, for any $\epsilon > 0$ there is a mixed strategy which assures P, a MOP firing hit with probability $\geq V - \epsilon$, described as a near optimal strategy, an $\epsilon$ strategy, where P can attain at most V.

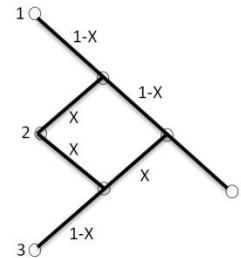

Figure 1. TOSG Protocol Game Tree Geometric Cockpit   Software Structure

III. TECHNICAL RISKS AND PROGRAMMATIC RISKS FOR DECISION OPIMIZATION

Key technical risks are developed for the constraints in a TOSG Protocol equation realizing multiple sensors detecting voids in the target communicating with a Simulation Facility. Risk is defined as the chance that a particular decision or action can give rise to a variety of outcomes for which the mathematical probability can be calculated [3,13]. Therefore, the programmatic risk is the integration of the Decision Optimization solution containing the constraint values. Risk is defined economically by the following equation[13],

$$RISK = T \times V \times C \qquad (6)$$

T = threat, the frequency of potentially adverse events and protection of goals, V = vulnerability, the likelihood of success of a particular organization, C = cost is the total impact of a particular threat exercised by a vulnerable target.

The mathematical probabilistic mitigating risk equation is the following:

$$RISK = (Pa)(1- Pe)(Ce) \qquad (7)$$

Pa = the probability of attack from the analysis of threat based on intelligence of the threat, current security environment[13] and other information to arrive at some indications of an event at worst case = 1.0. Use a value for likelihood of attack Pa other than the assumed worst case value of 1.0 to be used to help discriminate among the target set. Pe = system effectiveness is the product of Pi and Pn. Pi is the probability of interruption indicating how effective the protective system is in interrupting an adversary attack,  Pn = the probability of neutralization, how well response measures do in force-on-force conflicts with the adversary given interruption. Ce =

consequence of an event including prioritized targets. The risk of surface damage from the detonation of the MOP in the tunnel is discussed in the research on the Defense Cover[16,17] and the Side Effect Risks [22].

## IV. INVARIANT IMBEDDING OF TOSG PROTOCOL EQUATION WITHIN GAMES OF TIMING KERNEL

The invariant imbedding[7] of the TOSG equation (1) within the Games of Timing Optimal Strategy and Optimal Timing Interval Kernel equation [23] is crucial to the final Optimal Decision (TOSG) Protocol equation. The Optimal Strategy Timing Interval enables the Optimal Decision with the computations of parameters correlated to the TOSG Protocol Equation constraint solutions for (1) perceived threat, (2) threat missile, and (3) ground asset being attacked. Denman[7] in the research on invariant imbedding and optimal control stresses the "concept of optimizing the performance, yield or profit of a system." The optimization equation that Denman utilizes is a Linear Regulatory control problem equation. This optimization equation, i.e., Lagrangian, uses the invariant imbedding concept to completely define the performance of the Linear Regulator control system.

Also, a test for the integration error is made by Denman's two resulting equations from a coupled set of first-order differential equations leads to a two point boundary problem. The use of the invariant imbedding concept contains the TOSG Protocol equation within the optimization solution for the three constraints into the Kernel of the Games of Timing Optimal Strategy and Optimal Timing Interval. The Games of Timing Theory[26] definition of the Optimal Strategy and the Optimal Timing Interval states they are obtained as the solution of a certain integral equation with a positive kernel. In a wide category of cases this integral equation is equivalent to a certain linear differential equation or a system of linear first order differential equations. The Optimal Strategy will be obtained as the solution of a certain integral equation with a positive kernel[7].

## V. GAMES OF TIMING OPTIMAL STRATEGY AND OPTIMAL TIMING INTERVAL (MODEL 3)

### 1. Definition of Symmetric Game of Timing

The Symmetric Game of Timing [26] is a continuous game involving the Bilinear functional

$$\int_0^1 \int_0^1 K(x,y) dF(x) dG(y), K(x,y) = -K(y,x) \quad (8)$$

### 2. OPTIMUM PURE STRATEGY DEFINITION

For $x < y$, $K(x,y)$ is a strictly increasing function of x and a strictly decreasing function of y [26]. If $K(1^-,1) \leq 0$, there is an optimal pure strategy at 1; if $K(0,1) \geq 0$, there is an optimal pure strategy at 0. It will be proved there is a unique optimal strategy which is either a density from some point a to 1, or is a jump at 0 and a density from a to 1. If the quantity $K(x^-,y)$ varies in sign as y varies, let b be the value such that $K(b^-,b) = 0$ while $K(y^-,y) > 0$ for $b < y \leq 1$. The optimal strategy y is a density from a to 1 where $a > b$. It is shown that the determination of the density function depends on the solution of a certain integral equation with positive kernel, and the theory of such integral equations. It is shown for a general category of cases the optimal strategy can be obtained in terms of a system of ordinary linear differential equations. The proof of the uniqueness of an optimal strategy can be given in the following simpler form. If there are two optimal strategies, they must have the same spectrum.

### 3. SYMMETRIC CONTINUOUS GAMES DEFINITION

The equations considered are a class of Symmetric Continuous Games involving the bilinear function[26],

$$\int_0^1 \int_0^1 K(x,y) dF(x) dG(y) \quad (9)$$

where x,y range over the real numbers from 0 to 1 inclusive, the symmetry of the game reflecting itself in the skew symmetry of the kernel K(x,y) [26],

$$K(x,y) = -K(y,x) \quad (10)$$

Concerning the kernel K(x,y), suppose that for $x < y$, K(x,y) is a strictly increasing function of x and a strictly decreasing function of y. This property holds for $x > y$ by virtue of the skew-symmetry of K(x,y). Across the main diagonal $x = y$ this property may cease, i.e., there may be a jump of K(x,y), and $K(a+\delta, a)$ may be smaller than $K(a-\delta, a)$ for small positive $\delta$'s.

### 4. FORMAL DEFINITION OF A GAME OF TIMING

Such a Game is a Game of Timing by virtue of the following interpretation. The variables x and y may represent the times at which players I and II take certain specific actions; and it is profitable for each player to delay action as long as possible, provided its action is prior to its opponent's action. If the time x, y at which players I,II take action are near each other, there is a decided difference in the outcome accordingly as $x < y$ or $x > y$. Each player is thus subject to the following motive: it wishes to delay action so as to increase its reward, but at the same time not to delay so long that its opponent can with effectiveness precede it.

### 5. SYMMETRIC GAME OF TIMING

A Game will be a symmetric game of timing[19] if the kernel K(x,y) satisfies the following conditions:

$$K(x,y) = \begin{cases} A(x,y) & \text{for } x < y \\ 0 & \text{for } x = y \\ -A(x,y) & \text{for } x > y \end{cases}$$

where A(x,y) is continuous in $x \leq y$.

A(x,y) is a strictly increasing function of x and a strictly decreasing function of y.

A(x,y) has continuous first derivative in $x \leq y$ and the set of points where $A_x(x,y) = 0$ or $A_y(x,y) = 0$ contains no linear intervals, x = constant.

$\beta_1 < y < \beta_2$ or y = constant, $\alpha_1 < x < \alpha_2$

$A_x(x,y) \geq 0, \quad A_y(x,y) \leq 0 \quad \text{for } x \leq y$

### 6. OPTIMAL STRATEGY OF A GAME OF TIMING DEFINITION

The condition for A(x,y) makes a limit on places where either of these derivatives are zero. The optimal strategy of a game of timing is unique and consists either of (1) a density function from some point a to 1 or, (2) consists of a jump at 0 and a density from some point a to 1 [26].

7. SOLUTION OF A GAME OF TIMING WITH AN INTEGRAL EQUATION

The optimal strategy[7] will be obtained as the solution of a certain integral equation with a positive kernel. In a wide category of cases this integral equation is equivalent to a certain linear differential equation or a system of linear first order differential equations[7].

8. GENERAL CONDITIONS ON AN OPTIMUM STRATEGY FOR GAMES OF TIMING

$$\begin{cases} \text{If } A(1,1) \leq 0 \quad \text{a pure strategy at 1 is the unique optimum strategy} \\ \text{If } A(0,1) \geq 0 \quad \text{a pure strategy at 0 is the unique optimum strategy} \end{cases}$$

Suppose, $A(0,1) < 0$, $A(1,1) > 0$, and that there is an optimal strategy $F(x)$ for the game, and derive necessary conditions satisfied by $F(x)$. Then, [26]

$$V(y) = \int_0^1 K(x,y) \, dF(x) \geq 0 \quad \text{for all } y \qquad (11)$$

while,

$$\int_0^1 V(y) \, dF(y) = \int_0^1 \int_0^1 K(x,y) \, dF(x) dF(y) = 0$$

By the skew-symmetry of $K(x,y)$.      (12)

9. THE GAMES OF TIMING SOLUTION FOR TWO OPTIMAL STRATEGIES AND OPTIMAL STRATEGY TIMING INTERVAL EXISTS IF THEY HAVE THE SAME SPECTRUM.

The spectrum of $F(x)$ lies completely in the basic interval $b \leq x \leq 1$. This makes an assertion only if $b > 0$. A new game exists where the pay-off is $K(x,y)$, but where $x,y$ are limited to the interval $b \leq x \leq 1$, $b \leq y \leq 1$. This game of timing has a solution. Assuming the solution to this game will be $\varphi(x)$, $b \leq x \leq 1$, with $\varphi(1) = 1$, $\varphi(b) = 0$. Then extend $\varphi(x)$ below b by setting $\varphi(x) = 0$, for $x < b$. This contradiction establishes that only the game must be considered over the basic interval $b \leq x \leq 1$. Therefore only the basic interval, the timing interval will be considered for this basic interval. This basic interval will be the interval from 0 to 1, so that it is stated,

$$A(x,y) > 0 \quad \text{for } 0 < x \leq 1 \qquad (13)$$

Thus, it has been proven that if there are two optimal strategies, they must both have the same spectrum.

VI. TOSG PROTOCOL DECISION OPTIMIZATION EQUATION

The basic concept required is for the MOP bomber to make a decision to allocate sensor and weapon systems to threat launch events. The constraints on this decision are (1) perceived threat inventory, (2) threat missile, and (3) ground asset being attacked. Therefore the TOSG Protocol equation to Allocate Sensor and Weapon Systems Decisions[21] as the Objective Function, which will be invariant imbedded[7] within the kernel equation for the Games of Timing theory to obtain the Optimal Strategy time interval is stated as:

$$TOSG = ASWR + [\alpha[PTI-PTIc] + \beta[TM-TMc] + \gamma[GAA-GAAc]] \qquad (14)$$

ASWD = Allocate Sensor And Weapon Systems to Threat Launch Events Objective Function
PTI = Perceived Threat Inventory Constraint,
PTIc = PTI constraint value with risk
TM = Threat Missile Constraint,
TMc = TM constraint value with risk
GAA = Ground Asset Being Attacked,
GAAc = GAA constraint value with risk
$\alpha$ = PTI Lagrange Multiplier, $\beta$ = TM Lagrange Multiplier
$\gamma$ = GAA Lagrange Multiplier

The Lagrangian Optimization of the TOSG equation with objective function ASWR for each of the three constraints will be mathematically derived by obtaining the partial derivative of TOSG with respect to each of the three constraints and equating the expression to zero to enable a Lagrange Multiplier solution for each of the constraints. The risk equations[3,13] are included within the three TOSG Protocol optimization equation constraint equations. Then, a solution for the TOSG Protocol equation containing the constraint values can be achieved. The invariant imbedding[7] of equation (14) within the Games of Timing Optimal Strategy and Optimal Timing Interval Kernel equation is crucial to the final Decision Optimization equation. The Optimal Strategy Timing Interval enables the Optimal TOSG Decision with the computation of parameters correlated to the constraint solutions for (1) perceived threat, (2) threat missile, and (3) ground asset being attacked. Refer to Figure 2 for the Optimal TOSG Protocol Decision Equation Flow Chart. Figure 3 illustrates the theoretical Performance Analysis of the Three TOSG Protocol Models.

VI. SUMMARY

The TOSG Protocol Cockpit Software is theoretically developed for optimal control performance of the MOP bomber. The TOSG Protocol is composed of tactical game theory as a Zero Sum Two Person game and its interconnection to optimal strategy, aiming and evasion and games of timing theory. Three topological models with their correlated game theory basis are included in the TOSG optimization equation with risk constraints. The TOSG Protocol Game Tree Geometric Cockpit Software Structure represents the transit of the nuclear devices in the tunnel and the MOP bomber activity for the three models.

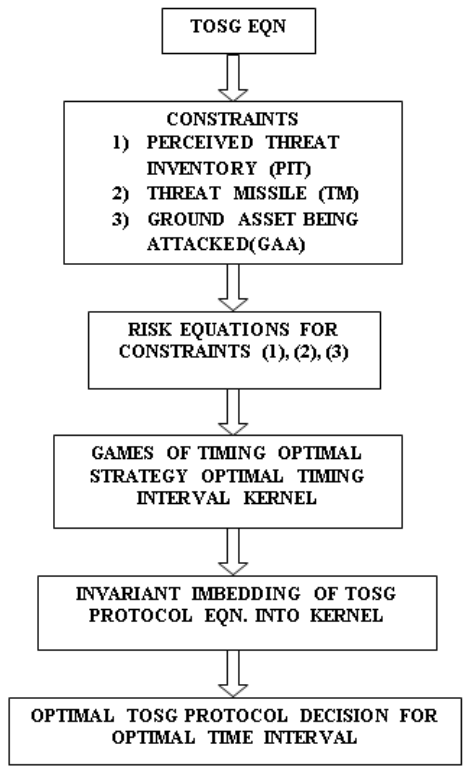

Figure 2. Optimal TOSG Protocol Decision Equation Flow Chart

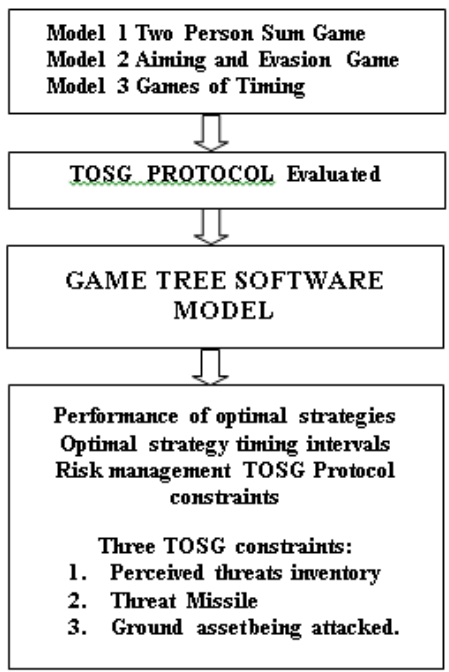

Figure 3. Performance Analysis Of TOSG Protocol For Three Models